\documentclass{article}
\usepackage{ltwol2e}
\def \ts{\thinspace}
\def \ns{\enspace}
\def \thetas{\theta^{*}}
\def \GeV{{\rm \enspace GeV}}
\def \TeV{{\rm \enspace TeV}}
\def \beq{\begin{equation}}
\def \eeq{\end{equation}}
\def \beqa{\begin{eqnarray}}
\def \eeqa{\end{eqnarray}}

\def \Nlike{N_{\|}}
\def \Noppo{N_{\times}}
\def \thetas{\theta^{*}}
 
\begin{document}


\title{Spin Issues in $t\bar{t}$ Production and Decay}

\author{Gregory Mahlon$^{*}$}

\address{Department of Physics, McGill University, 
3600 University St., Montr\'eal, QC  H3A 2T8, Canada\\
Electronic address:  mahlon@physics.mcgill.ca}

\twocolumn[\maketitle\abstracts{
We describe the off-diagonal spin basis for observing angular
correlations in top quark pair production events at the Fermilab 
Tevatron.  For events initiated by quark-antiquark annihilation,
the top and antitop quark spins are 100\% correlated in this basis:  
a spin-up top quark is always accompanied by a spin-down antitop 
quark and vice versa.  Inclusion of the gluon-gluon initial state
lowers the fraction of unlike spin events to 92\%.
Nevertheless, the angular correlations between the top
and antitop quark decay products are twice as large 
in the off-diagonal basis as those in the more traditional
helicity basis.  We give two brief examples of how the presence
of new physics would alter these correlations.
\hfill McGill/98-34
}]


Until the discovery of the top quark, most studies of spin
in high energy physics were formulated in terms of the
helicity basis.  For ultrarelativistic particles, this 
is appropriate.  However, in general, the direction and degree
of polarization of a massive spinning particle depends on
how it was produced.  Thus, for moderate particle energies,
it should not be surprising to find that the optimal axis
for studying spin correlations is something other than 
the particle's direction of motion.

\begin{figure}[h]
\vskip6.0cm
\includegraphics{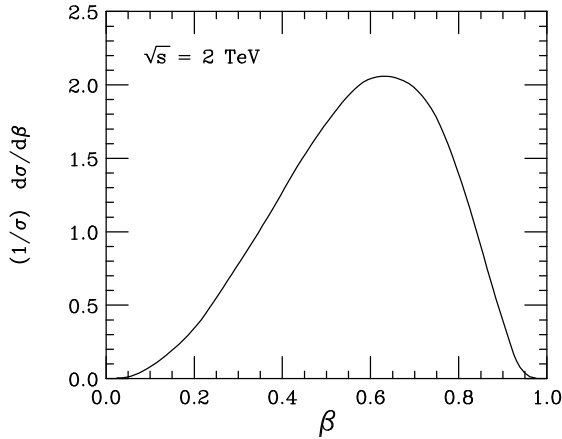}
\caption[]{Differential cross section for $t\bar{t}$
production as a function of the zero momentum frame speed $\beta$
of the top quark for the Tevatron 
with $\protect\sqrt{s}=2\TeV$.\protect\cite{PDFs}}
\label{beta}
\end{figure}

In this talk, we will discuss the spin correlations in
top quark pair ($t\bar{t}$\ts) production at the 
Tevatron with a center-of-mass energy of 2 TeV.
The top quarks in these events are only moderately relativistic:
they typically have speeds of $\beta\sim 0.6$
in the zero momentum frame (ZMF) of the
initial partons (see Fig.~\ref{beta}).  
Therefore, it is not surprising to learn that the optimal
basis for studying the spin correlations in the $t\bar{t}$
system at this energy
is not the helicity basis, but rather the off-diagonal
basis.\cite{ODbasis}  As we shall see, the spin correlations
in the off-diagonal basis are a factor of two larger than the
correlations in the helicity basis.

Before examining $t\bar{t}$ production, let us review
a few facts about top quark decay.
Because of the enormous width of the top quark
($\Gamma_t = 1.6 \GeV$ in the Standard Model), 
its decay occurs before either hadronization 
(governed by the scale $\Lambda_{\rm QCD}$) or depolarization
(governed by the scale $\Lambda_{\rm QCD}^2/M_t$) can take 
place.\cite{Bigi} The dominant Standard Model decay
chain is
\beq
\includegraphics{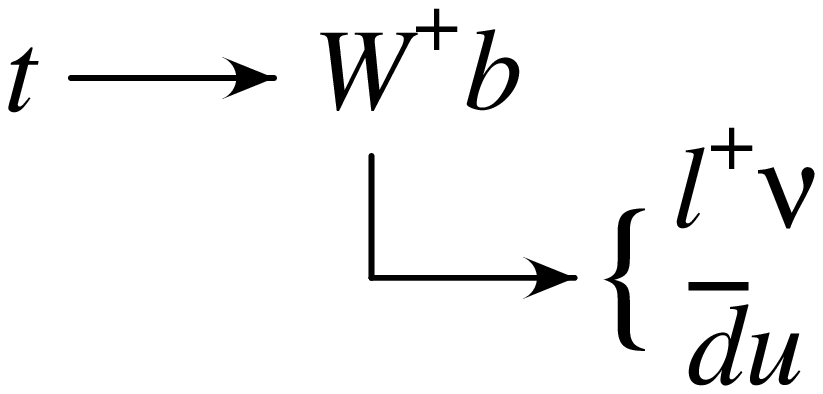}
\phantom{\Bigl[\over\Bigl[}
\eeq
For concreteness, we will describe the leptonic $W$ decay.
However, everything which we say about the charged lepton
applies equally to the $d$-type quark in a hadronic decay.

\begin{figure}[t]
\vskip4.2cm
\includegraphics{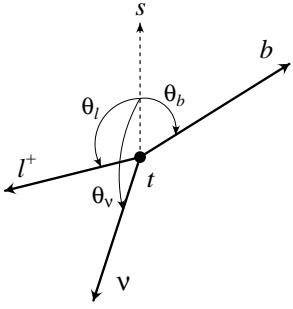}
\caption[]{Definition of the top quark decay angles in the
top quark rest frame.  The direction of the top quark spin
is indicated by the vector $s$.  Although we have drawn this
figure assuming a leptonic $W$ decay, the same correlations
hold in a hadronic decay if we replace the charged lepton by
the $d$-type quark and the neutral lepton by the $u$-type
quark.}
\label{DecayAngles}
\end{figure}

\begin{figure}[t]
\vskip5.8cm
\includegraphics{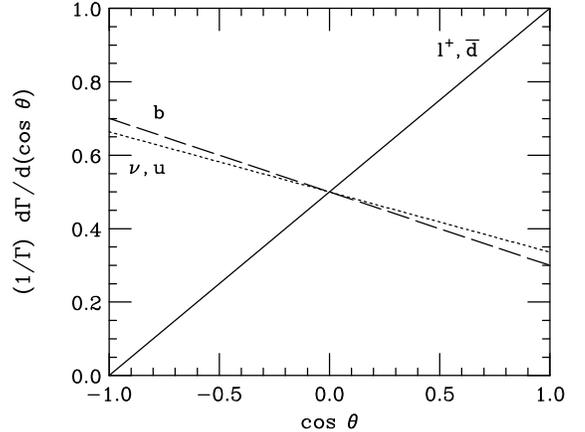}
\caption[]{Angular correlations in the decay of a spin-up top
quark.  The lines labeled $\ell^{+}$, $\bar{d}$, $b$, $\nu$,
and $u$ are the angle between the spin axis and the particle
in the rest frame of the top quark.}
\label{dGammaPlot}
\end{figure}

We define the decay angles in the top quark rest frame
with respect to top quark spin vector $s$,
as shown in Fig.~\ref{DecayAngles}.
The decay angular distributions of a spin-up top quark
are simply linear
in the cosine of these decay angles:
\beq
{1\over\Gamma}\thinspace
{ {d\Gamma}\over{d(\cos\theta_i)} }
=
{1\over2}
\Bigl( 1+\alpha_i\cos\theta_i \Bigr),
\label{dGamma}
\eeq
where $\theta_i$ is the decay 
angle of the $i$th decay product.\cite{alphas}
The distribution for
spin-down top quarks has a minus sign in front
of the $\cos\theta_i$ term.
The degree to which each decay product is correlated
with the spin is encoded in the value of $\alpha_i$ 
(see Table~\ref{alphaTable}).
Notice that the charged lepton (or $d$-type quark) is
maximally correlated, with $\alpha_i=1$.
Thus, the most distinctive distribution plots the angle between
the spin axis and
the charged lepton in the top quark rest frame 
(see Fig.~\ref{dGammaPlot}).

When we write the decay matrix element in an arbitrary
Lorentz frame, we find that
the natural 4-vectors are not
the top quark momentum $t$ and its spin 
vector $s$ (normalized such that $s_\mu s^\mu = -1$).
Instead, it is more convenient to use the combinations
\beq
t_1 \equiv \hbox{$1\over2$} (t+ms) \quad{\rm and}\quad
t_2 \equiv \hbox{$1\over2$} (t-ms),
\label{breakdown}
\eeq
where $m$ is the mass of the top quark.  
In the top quark rest frame, the spatial parts of $t_1$ and $s$
point in the same direction, since in this frame $t = (m,{\bf 0})$.
In some other frame, however, these vectors
are not parallel.\footnote{This follows trivially from the observation
that $t_1$ is a massless vector, whereas $s$ has been constructed 
to be spacelike.}
In this case, the form of the
matrix element clearly indicates
that the preferred charged lepton emission axis is the
spatial part of $t_1$.
Hence, we regard $t_1$ as the
appropriate generalization of the spin axis to an arbitrary
reference frame.  

\begin{table}[t]
\centering%
\caption{Correlation coefficients $\alpha_i$ for both
semileptonic and hadronic top quark decays.  The first
two entries are a function of $M_t^2/M_W^2$, and have
been evaluated for  
$M_t = 173.8 \GeV$ and $M_W = 80.41 \GeV$.\protect\cite{PDG,BMLbasis}
\lower2pt\hbox{\protect\phantom{j}}
\label{alphaTable}}
\begin{tabular}{ccccc}
\multispan5\hrulefill \\[0.05cm]
&Decay Product &\quad& $\alpha_i$ & \\[0.2cm]
\multispan5\hrulefill \\[0.1cm]
\qquad\qquad\quad\thinspace\enspace
& $b$  &\quad&  $-0.40$  & \\
& $\nu_{\ell}, u,$ or $c$
     &\quad& $-0.33$ & \thinspace\enspace\quad\qquad\qquad \\
& $\bar{\ell}, \bar{d},$ or $\bar{s}$
     &\quad& $\protect\phantom{-}1.00$ & \qquad\quad \\[0.1cm]
\multispan5\hrulefill 
\end{tabular}
\end{table}

Unless there is some type of spin asymmetry in the data,
the opposite dependence upon $\cos\theta_i$ for spin-up
and spin-down top quarks will wash out the correlations,
leaving a flat distribution.
Considered individually, the top quarks in $t\bar{t}$ pairs at the
Tevatron are essentially unpolarized:\footnote{The
small QCD loop-induced transverse polarization of the top quarks
may be ignored for our purposes.} spin-up and spin-down top
quarks are produced in equal numbers.  
However, there is an asymmetry when we examine the top
and antitop quarks as a pair.
In general, the number of pairs
where both quarks have spin up or spin down ($\Nlike$)
is {\it not}\ equal to the number of pairs where one
quark is spin up and the other is spin down ($\Noppo$).
In this situation, 
correlations are visible in a joint distribution
containing one decay angle from the top side of the
event and one decay angle 
from the antitop side of the event.  Denoting these two decay
angles by $\theta_i$ and 
${\bar\theta}_{\bar\imath}$ respectively, we have
\beq
{1\over\sigma}\thinspace 
{ 
{d^2\sigma}
\over
{ d(\cos\theta_i) d(\cos\bar\theta_{\bar\imath}) }
} = 
{1\over4}
\biggl[
1 + 
{
{ N_{\|} {-} N_{\times} }
\over
{ N_{\|} {+} N_{\times} }
}
\thinspace
\alpha_i \bar\alpha_{\bar\imath}
\cos\theta_i \cos\bar\theta_{\bar\imath} 
\biggr]
\label{Double}
\eeq
for the complete production and decay process,
$p\bar{p}\rightarrow t\bar{t} \rightarrow \hbox{6-body final state}$.
Eq.~(\ref{Double}) explicitly exhibits the dependence of
the correlations on the production and decay stages
of the event.  Production is represented by the pairwise
spin asymmetry $(\Nlike{-}\Noppo)/(\Nlike{+}\Noppo)$.
This factor depends upon the choice of spin basis and
may be maximized by employing the off-diagonal
basis\ts\cite{ODbasis} (see below).
Decay is represented by the correlation
coefficients $\alpha_i$ and ${\bar\alpha}_{\bar\imath}$ 
as well as the decay angles
$\theta_i$ (measured in the $t$ rest frame) 
and ${\bar\theta}_{\bar\imath}$
(measured in the $\bar{t}$ rest 
frame).\footnote{The experimental challenge of reconstructing
the $t$ and $\bar{t}$ rest frames with sufficient accuracy
is one which must be met no matter what spin basis is employed.}
Our choice of which decay angles we measure determines how
well we can see a given production asymmetry.
From this point of view, we want to make the $\alpha$'s
as large as possible --
{\it i.e.}\ we should choose to measure
the decay angles of the charged leptons.
Because 2-dimensional distributions generally require high 
statistics for accurate mapping, it may be desirable
to construct 1-dimensional or even 0-dimensional
projections of Eq.~(\ref{Double}).  Refs.~\ref{BMLbasis}
and~\ref{Willenbrock} contain some suggestions on how to do this.

\begin{figure}[t]
\vskip3.25cm
\includegraphics{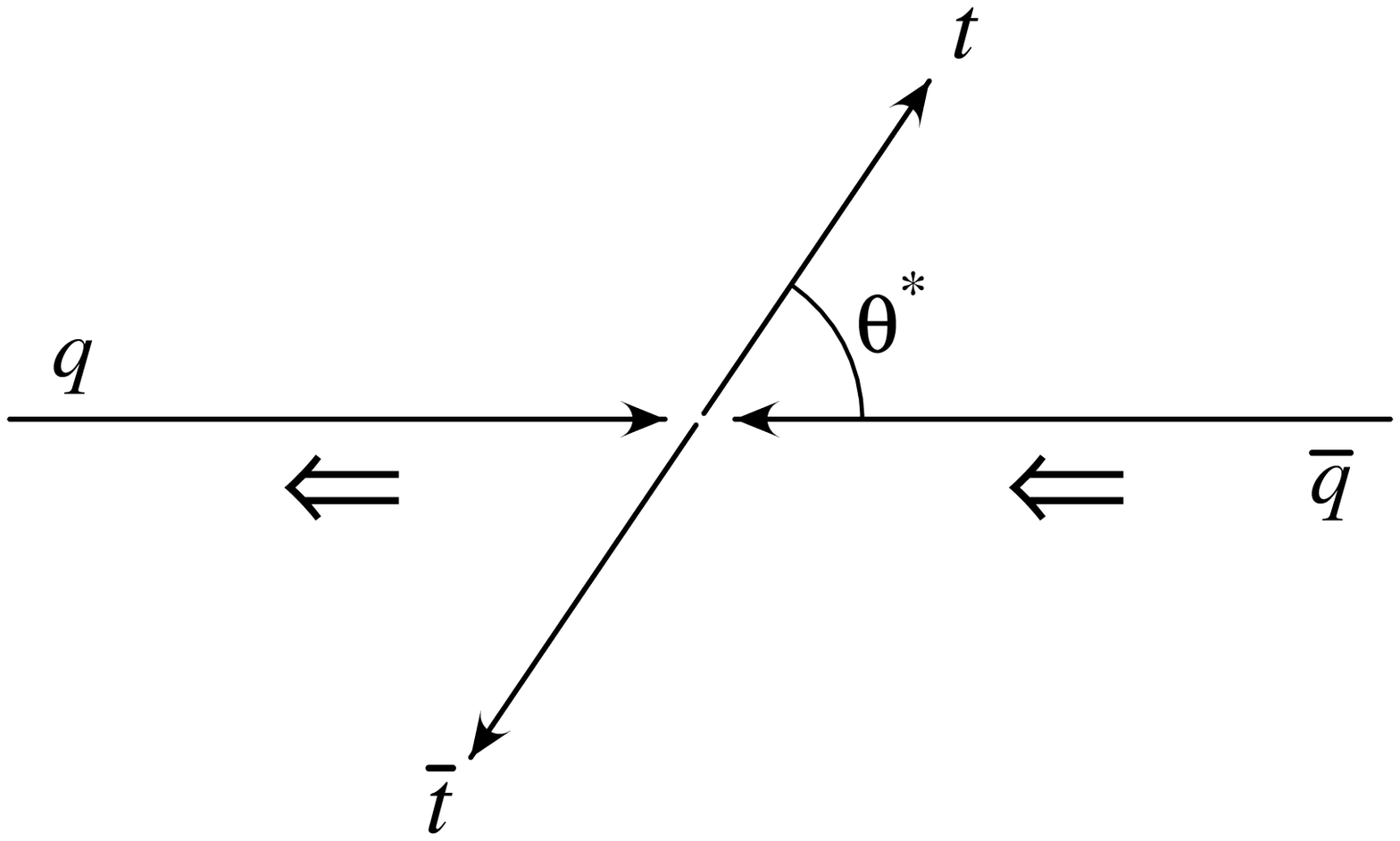}
\caption[]{The process $q\bar{q}\rightarrow t\bar{t}$, viewed
in the zero momentum frame.  The initial $q$ and $\bar{q}$
must have opposite helicities to couple to the gluon
in the intermediate state.  One of the two permitted $q\bar{q}$ spin
configurations is indicated by the wide arrows.
}
\label{thresh}
\vskip5.0cm
\includegraphics{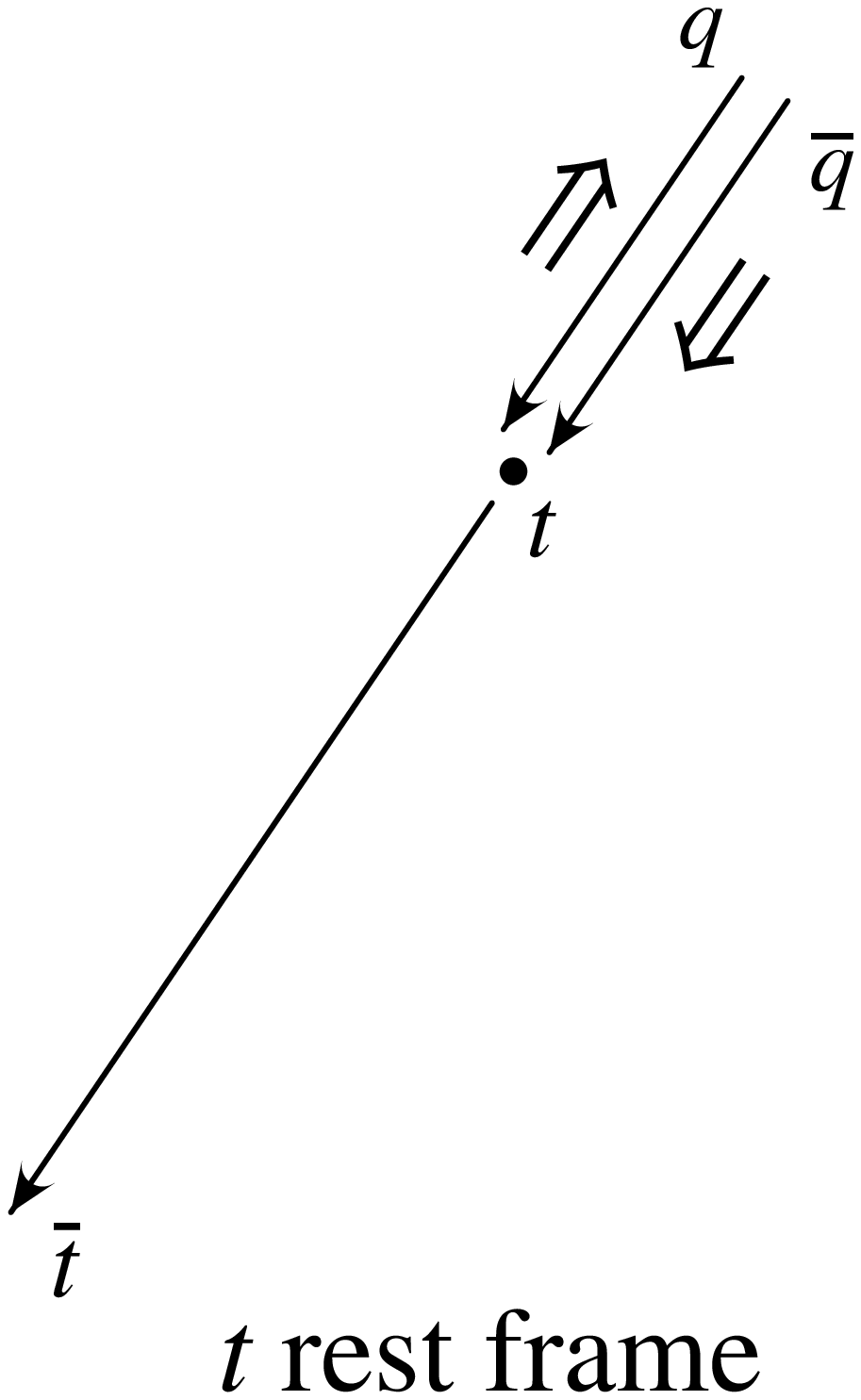}
\includegraphics{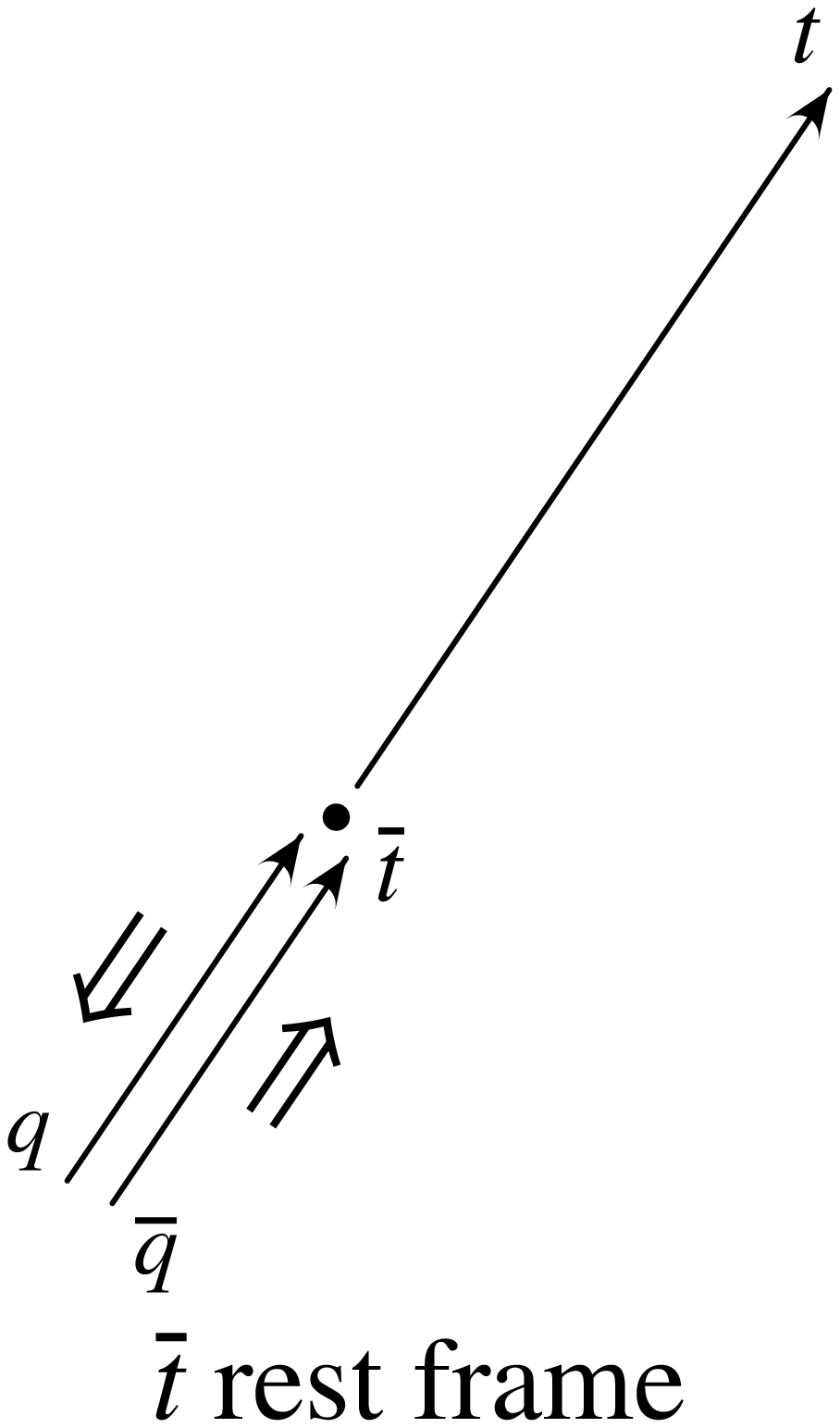}
\caption[]{The process $q\bar{q}\rightarrow t\bar{t}$
in the ultrarelativistic limit, viewed in the $t$ and $\bar{t}$
rest frames.  The very large boost factor forces the initial
$q\bar{q}$ pair to be aligned with the $\bar{t}$ in
the $t$ rest frame.  The wide arrows indicate
one of the two permitted $q\bar{q}$ spin configurations.
}
\label{ultra}
\end{figure}

We now consider how spin issues relate to $t\bar{t}$ production.
Because the majority
($\sim 90\%$) of the cross section comes from the quark-antiquark
initial state, we will first focus our attention on the process
$q\bar{q}\rightarrow t\bar{t}$, as illustrated in Fig.~\ref{thresh}.
We describe this event in terms of the ZMF
production angle $\thetas$ 
and the ZMF speed of the top quark $\beta$.
The initial quarks are firmly in the ultrarelativistic regime 
since $m_t \gg m_q$.  Because
the $q\bar{q}g$ coupling in QCD
is helicity-conserving, we conclude that the initial $q$ and $\bar{q}$
have unlike helicities.   

Suppose that the $t\bar{t}$ pair in Fig.~\ref{thresh} 
is produced very near threshold.
Then, the $t$ rest frame and the $\bar{t}$
rest frame both coincide with the ZMF to a good approximation.
Knowledge of the $q$ and $\bar{q}$ helicities translates into
knowledge of the total angular momentum along the beam axis:
{\it i.e.}\ the unlike $q$ and $\bar{q}$ helicities implies
unlike $t$ and $\bar{t}$ spins measured along the beam axis.
Along any other axis,
there will be a superposition of like
and unlike $t$ and $\bar{t}$ spins.   Thus, at threshold, 
the helicity basis does not describe the physics most simply.

On the other hand, if the $t\bar{t}$ pair is produced 
in the ultrarelativistic regime far above threshold,
then the picture in the $t$ and $\bar{t}$ rest frames is
vastly different from the picture in the ZMF (see
Fig.~\ref{ultra}).  
In the rest frame of either top, the momenta of the
other top and both light quarks are essentially parallel.
The light quarks still have opposite helicities.
Knowledge of the $q$ and $\bar{q}$ helicities thus translates
into knowledge of the $t$ and $\bar{t}$ helicities.
Using any other spin axis, there would be a superposition
of like and unlike spins.  Hence, we recover the 
rationale for employing the helicity basis to describe
ultrarelativistic fermions.

\begin{figure}[t]
\vskip4.2cm
\includegraphics{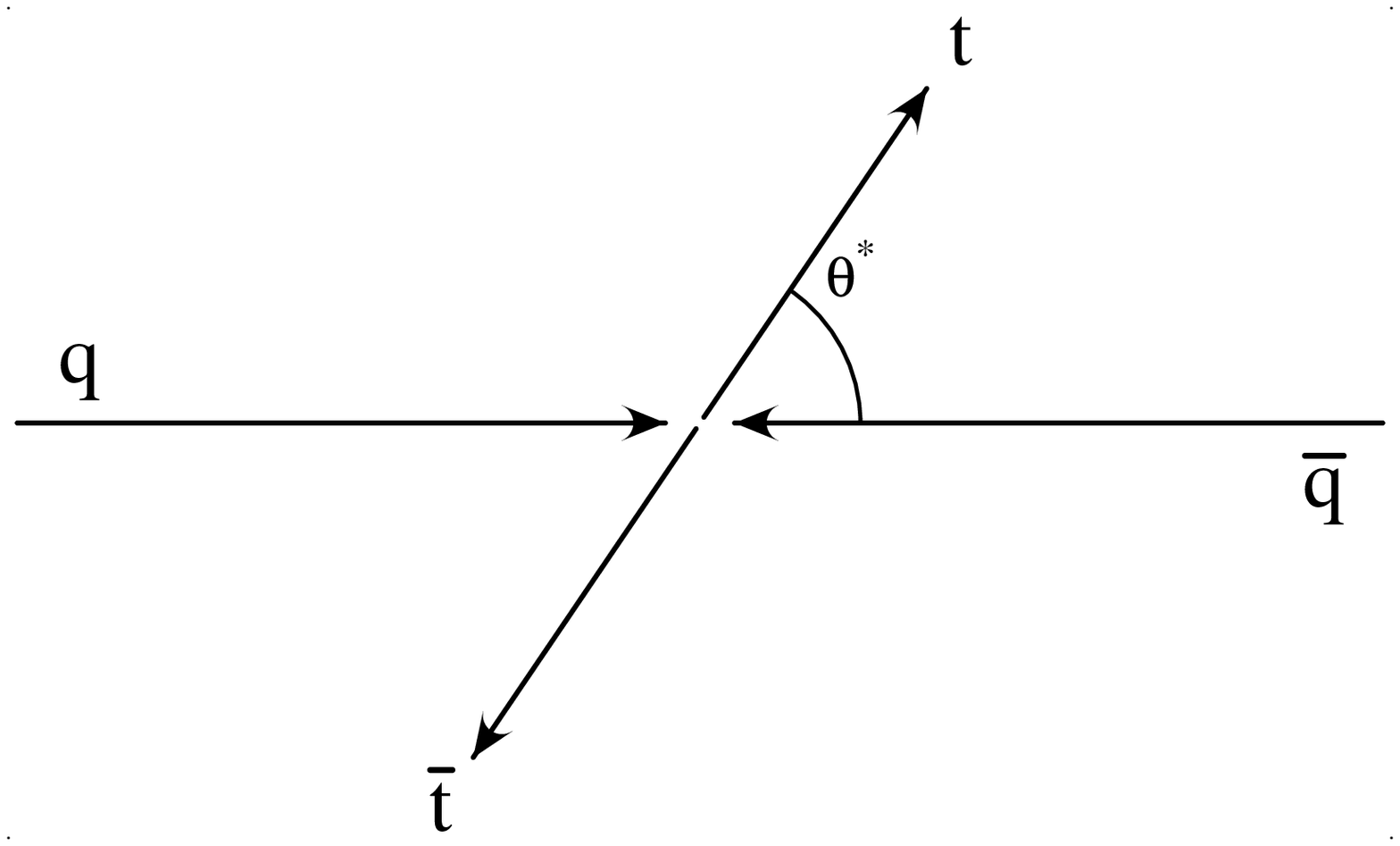}
\includegraphics{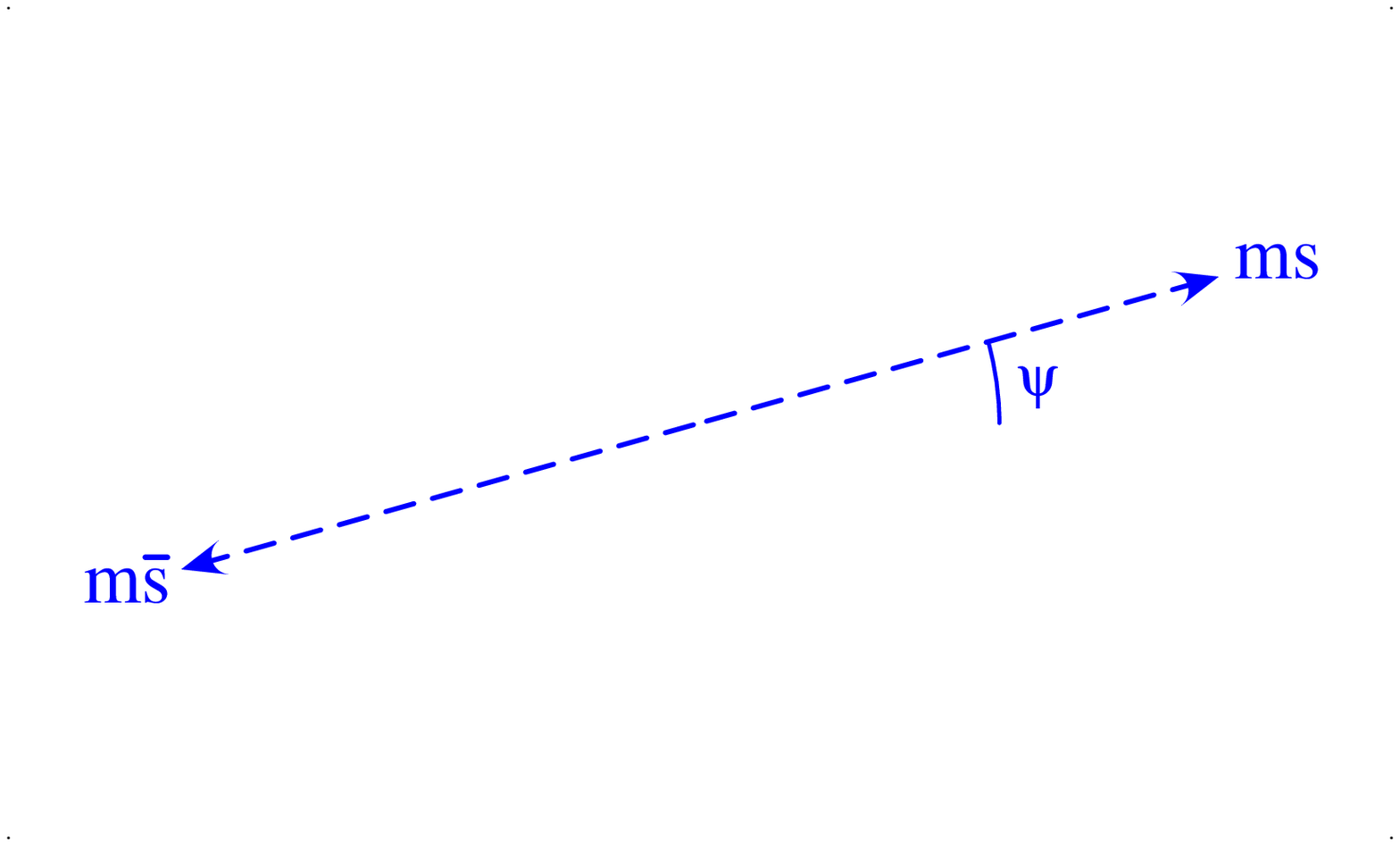}
\includegraphics{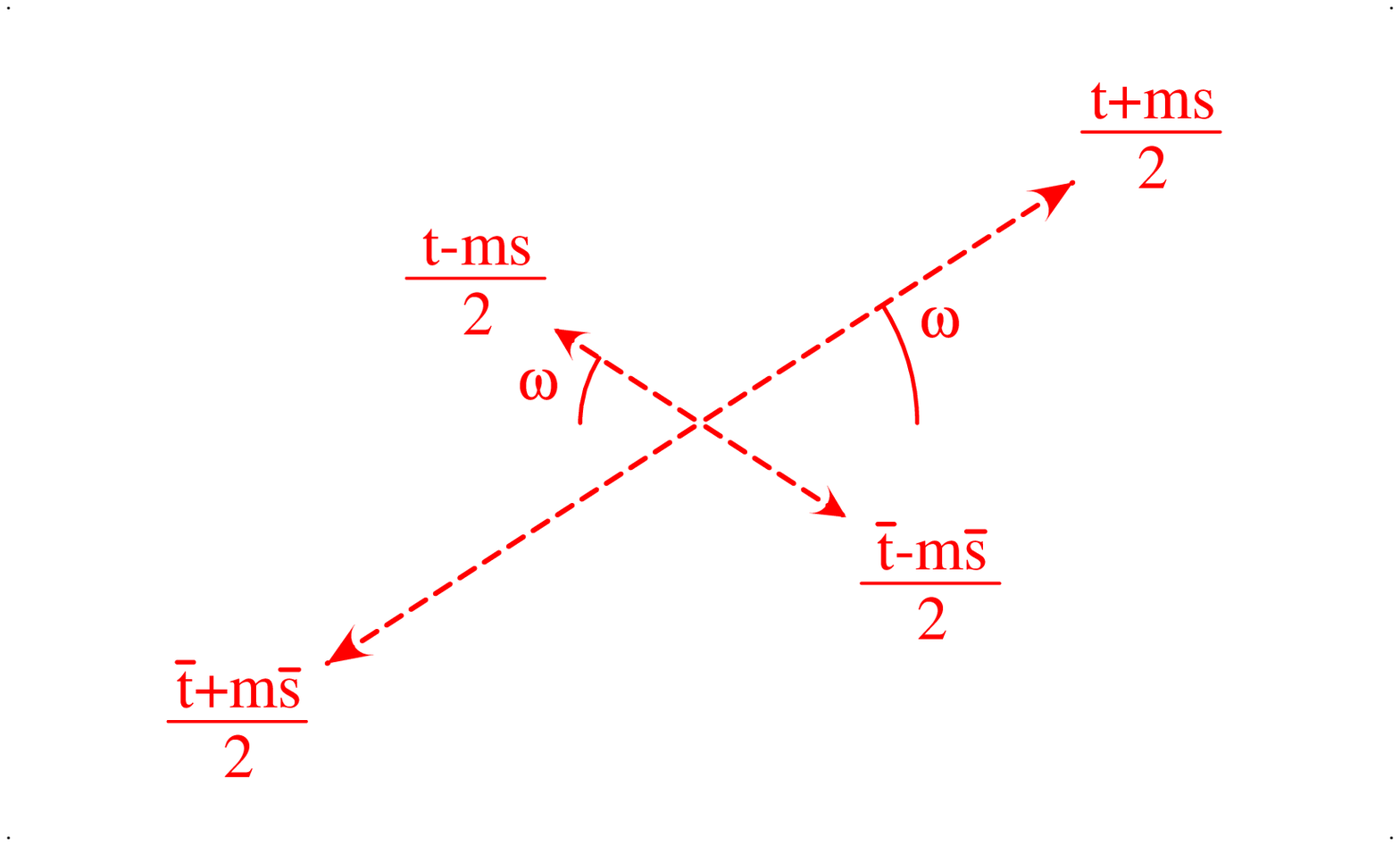}
\caption[]{Anatomy of a $q\bar{q}\rightarrow t\bar{t}$ event
in the zero momentum frame.   All vectors lie in the production
plane.  
The top quark is produced at an angle $\theta^*$ with
respect to the beam axis.  The off-diagonal basis spin
vector $s$ makes an angle $\psi$ (given by Eq.~(\protect\ref{PSI}))
with respect to the beam axis.  The vectors $(t\pm ms)/2$,
where $m$ is the top quark mass, indicate
the preferred emission directions for the charged lepton 
or $d$-type quark from the decaying $W^{+}$ 
(see Eq.~(\protect\ref{OMEGA})).  
The vectors describing the antitop
lie back-to-back with the corresponding top quark vectors.}
\label{Anatomy}
\end{figure}

Note that in both extremes ($\beta\rightarrow 0$
and $\beta\rightarrow 1$),
there is a basis in which the $t$ and $\bar{t}$ spins are 100\%
correlated:  a spin-up $t$ implies a spin-down $\bar{t}$ and
vice versa.  This suggests that we should seek a basis for
which this property holds for arbitrary $\beta$.
The authors of Ref.~\ref{ODbasis} have constructed 
such a basis, which they call
the {\it off-diagonal basis}.
This basis takes its name from the fact that for 
this choice of spin axis, the top pairs coming from
$q\bar{q} \rightarrow t \bar{t}$ are
purely in a state of unlike spins independent of 
their production angle and ZMF speed.
The important vectors in this basis are
illustrated in Fig.~\ref{Anatomy}.  
The direction of the spin vector $s$ in the off-diagonal basis
is given by the angle $\psi$, where
\beq
\tan\psi = 
{
{\beta^2 \cos\thetas\sin\thetas}
\over
{ 1-\beta^2\sin\thetas}
}.
\label{PSI}
\eeq
The vectors $t_1$ and $t_2$ 
({\it cf.}\ Eq.~(\ref{breakdown}))
have a much
simpler dependence on $\thetas$ and $\beta$:  they are at
an angle $\omega$ with respect to the beam, where
\beq
\sin\omega = \beta\sin\thetas.
\label{OMEGA}
\eeq
The $\beta\rightarrow 0 $ and $\beta\rightarrow 1$ limits
are particularly transparent in Eq.~(\ref{OMEGA}):
near threshold, $\omega\rightarrow 0$ (the beam direction)
while at very high energy, $\omega\rightarrow \thetas$ (the
helicity direction).
Either of~(\ref{PSI}) or~(\ref{OMEGA}) may be taken as the
relation defining the off-diagonal basis.  
In any case,
the vectors ${1\over2}(t\pm ms)$ and ${1\over2}(\bar{t}\pm m\bar{s})$
are special:  for the up-down spin configuration
the preferred emission directions of the charged leptons
are ${1\over2}(t+ms)$ for the $\ell^{+}$ and
${1\over2}(\bar{t}+m\bar{s})$ for the $\ell^{-}$.
For the down-up spin configuration, the charged leptons
prefer the directions ${1\over2}(t-ms)$ and
${1\over2}(\bar{t}-m\bar{s})$.

\begin{figure}[t]
\vskip6.2cm
\includegraphics{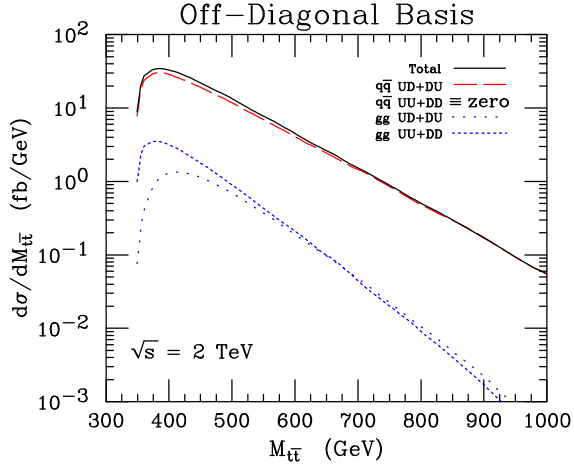}
\caption[]{Differential cross section for $t\bar{t}$ production
as a function of the $t\bar{t}$ invariant mass $M_{t\bar{t}}$
for the Tevatron with $\protect\sqrt{s} = 2.0 \TeV$, decomposed
into like and unlike spins of the $t\bar{t}$ pair using
the off-diagonal basis for both $q\bar{q}$ and $gg$ production
mechanisms.\protect\cite{PDFs}
}
\label{mttbar}
\end{figure}

If $q\bar{q}$ were the only initial state at the Tevatron,
then we would have $(\Nlike{-}\Noppo)/(\Nlike{+}\Noppo) = -1$.
However, approximately 10\% of the top pair events are initiated
by a $gg$ initial state.   The spin-1 nature of the gluon
translates into different $t\bar{t}$ correlations.  For example,
near threshold, the gluons must have like helicities to form
a $t\bar{t}$ final state, since opposite helicity gluons
would have a total spin projection of $\pm2$ along the beam
axis, whereas the maximum spin projection of the $t\bar{t}$
pair is $\pm1$.\footnote{There can be no orbital angular momentum
at threshold.}  Thus, low mass $t\bar{t}$ pairs tend to have
like spins along the beam direction.   At very high
energies, we recover the preference for opposite helicities
since in the top rest frame the gluons have parallel momenta.
The upshot of this is that the off-diagonal basis, which works
well for the $q\bar{q}$ initial state, is not ideally suited to the
$gg$ initial state.  Nevertheless, since $q\bar{q}$ 
dominates the total cross section, the off-diagonal basis is
still an excellent choice,
although the spin-pair asymmetry is degraded
a bit.  In Fig.~\ref{mttbar}
we have plotted the distribution in the $t\bar{t}$ invariant
pair mass, broken down into the contributions from $q\bar{q}$
and $gg$ to the like and unlike spin configurations.  Including
the gluons, we find that 92\% of the $t\bar{t}$ pairs have
unlike spins, corresponding to correlations which are about
a factor of two larger than those in the helicity basis
(see Table~\ref{SpinFractions}).

\begin{table}[t]
\centering
\caption{Dominant spin fractions and asymmetries for the helicity
and off-diagonal bases for top quark pair production 
at the Tevatron with $\protect\sqrt{s} = 2.0 \TeV$.
\lower2pt\hbox{\protect\phantom{j}}
\label{SpinFractions}}

\begin{tabular}{ccccc}
\multispan5\hrulefill \\[0.05cm]
&Basis        & Spin Content 
              & $\displaystyle{ { N_{\|} {-} N_{\times} }
                  \over { N_{\|} {+} N_{\times} } }$
              &  \\[0.3cm]
\multispan5\hrulefill \\[0.1cm]
\quad\ns\ns
&helicity     &   70\% unlike
              & $-0.39$ &\quad\ns\ns \\
&off-diagonal &   92\% unlike
              & $-0.84$ & \\[0.1cm]
\multispan5\hrulefill 
\end{tabular}
\end{table}

The presence of non-Standard Model physics would alter the
correlations we have just described.  
One interesting possibility
involves
a new scalar or pseudoscalar state which couples
strongly to the top quark.\cite{techni}  
Typically, this kind of new physics will manifest as a bump 
in the $t\bar{t}$ invariant mass spectrum.  
Suppose that we are fortunate and such a bump is observed.  
Then, an analysis
of the $t\bar{t}$ spin correlations can tell us about the
parity of the new state.  In particular, a pseudoscalar
coupled to $t\bar{t}$ produces top pairs which have like
spins {\it independent of the choice of spin basis}.
On the other hand, the top pairs formed via an intermediate
scalar would have like spins 100\% of the time only in the helicity
basis.  
In (almost) any other basis, such as the off-diagonal basis,
a scalar would contribute to both the like and unlike 
spin configurations.  The exception is the basis
where the scalar couples exclusively to unlike spin top pairs.
This basis is obtained by taking the spin vector to be at right
angles to the antitop direction of motion in the top rest frame.
In the lab frame, the corresponding direction of
${1\over2}(t+ms)$ is
\beq
\sin\omega =
\beta\sin\thetas + \sqrt{1{-}\beta^2} \cos\thetas
\label{Orthogacity}.
\eeq
Our point is that should a bump in the $M_{t\bar{t}}$ spectrum
be observed, it would be worthwhile to measure the spin correlations
for those events within the peak in more than one spin basis.
Doing so allows us to distinguish between scalar and
pseudoscalar intermediate states, something which could not
be done using only the helicity basis.

As a second example of how new physics would alter the
$t\bar{t}$ spin correlations,  suppose that there is a
charged Higgs decay of top, $t \rightarrow H^{+}b$.
Then, the value of $\alpha$ appearing in Eq.~(\ref{dGamma})
would be different.  In particular, we would have 
$\alpha_b = 1$ and
$\alpha_j = (-\xi^2 + 1 + 2\xi\ln\xi)/(\xi-1)^2$, where
$\xi\equiv M_t^2 / M_H^2$, and $j$ is either $H^{+}$ decay
product (independent of whether the $H^{+}$ goes to $c\bar{s}$ 
or $\tau^{+}\nu_\tau$).\cite{BMLbasis}  In a sample containing
both Standard Model and charged Higgs top decays, the observed
correlations would depend upon the relative size of the two
branching ratios.

In conclusion, the extremely short top lifetime provides
us with the opportunity to study the spin properties of
a ``free'' quark.  In general, the number of produced
like-spin top quark pairs is not equal to the number of
unlike-spin top pairs.  For the process $q\bar{q} \rightarrow t\bar{t}$,
100\% of the top quark pairs have unlike spins with respect
to the off-diagonal basis,\cite{ODbasis} defined in Eqs.~(\ref{PSI}) 
and~(\ref{OMEGA}).  When the $gg\rightarrow t\bar{t}$ process
is included, some like-spin top quark pairs are produced.
However, the spin correlations in the off-diagonal basis
are still twice as large as those in the helicity basis.
The largest correlations involve the charged leptons (or $d$-type
quarks) with respect to the spin axis in the rest frame
of the parent top quark.  These correlations can be 
significantly altered from their Standard Model values if the
top quark is strongly coupled to new physics.


\end{document}